\documentclass[11pt]{article}
\pdfoutput=1
\usepackage{jheppub}
\usepackage{amssymb,amsfonts}
\usepackage{mathrsfs}
\usepackage{epsfig}
\let\savenumberline\numberline
\def\numberline#1{\savenumberline{#1.}}
\makeatletter
\renewcommand{\@seccntformat}[1]{\csname the#1\endcsname.\,\,}
\makeatother

\newcommand{\R}{{\bf R}}

\newcommand{\CF}{{\cal F}}
\newcommand{\CG}{{\cal G}}

\newcommand{\CL}{{\cal L}}
\newcommand{\CM}{{\cal M}}
\newcommand{\CN}{{\cal N}}
\newcommand{\CO}{{\cal O}}

\newcommand{\CW}{{\cal W}}

\newcommand{\RD}{{{\mathrm D}}}
\newcommand{\barRD}{{\bar{\mathrm D}}}

\newcommand{\SF}{{\mathscr F}}
\newcommand{\SM}{{\mathscr M}}
\newcommand{\QB}{{Q^{\ }_{\textrm{BRST}}}}

\newcommand{\LS}{L}
\newcommand{\TS}{T}
\newcommand{\ES}{E}
\newcommand{\MS}{M}
\newcommand{\p}{\partial}
\renewcommand{\bar}[1]{\overline{#1}}

\newcommand{\be}{\begin{equation}}
\newcommand{\ee}{\end{equation}}
\newcommand{\bea}{\begin{eqnarray}}
\newcommand{\eea}{\end{eqnarray}}

\newcommand{\ie}{{\it i.e.}}
\newcommand{\eg}{{\it e.g.}}
\newcommand{\diff}{{\rm Diff}}

\makeatletter
\def\@fpheader{\relax}
\makeatother
\usepackage{graphicx}
\usepackage{latexsym}
\title{\ \vspace{1.5in} \\ The Geometry of Time in Topological Quantum Gravity\\ of the Ricci Flow}
\author{Alexander Frenkel${}^a$, Petr Ho\v{r}ava${}^{b,c}$ and Stephen Randall${}^{b,c}$}
\affiliation{${}^a$Stanford Institute for Theoretical Physics and Department of Physics\\
Stanford University, Stanford, CA, 94305-4060, USA\medskip\\
${}^b$Berkeley Center for Theoretical Physics and Department of Physics\\
University of California, Berkeley, CA, 94720-7300, USA\medskip\\
${}^c$Physics Division, Lawrence Berkeley National Laboratory\\
Berkeley, CA 94720-8162, USA}
\abstract{We continue the study of topological nonrelativistic quantum gravity associated with a family of Ricci flow equations on Riemannian manifolds.  This topological gravity is of the cohomological type, and it exhibits an ${\cal N}=2$ extended BRST symmetry.  In our previous work, we constructed this theory in a two-step procedure in the appropriate nonrelativistic ${\cal N}=2$ superspace, first presenting a topological theory of the spatial metric $g_{ij}$, and then adding the superspace versions of the lapse and shift variables $n$ and $n^i$ while gauging the symmetries of foliation-preserving spacetime diffeomorphisms.  In the relation to Perelman's theory of the Ricci flow, the role of Perelman's dilaton is played by our nonprojectable lapse.  Here we demonstrate that this construction is equivalent to a standard one-step BRST gauge-fixing of a theory whose fields are $g_{ij}$, $n^i$ and $n$, and whose gauge symmetries consist of (i) the topological deformations of $g_{ij}$, and (ii) the ultralocal nonrelativistic limit of spacetime diffeomorphisms.  The supercharge $Q$ of our superspace construction plays the role of the BRST charge.  The spacetime diffeomorphism symmetries appear in an interestingly ``shifted'' form, which may be of broader interest for nonrelativistic quantum gravity outside of the present topological context.  In contrast to the foliation-preserving spacetime diffeomorphisms, the gauge symmetries identified in this paper act nonprojectably on time, making it clear that this theory has no local propagating degrees of freedom.  We point out an intriguing dual interpretation of the same theory, as a gauge fixing of a dual copy of ultralocal spacetime diffeomorphisms, with the role of ghosts and antighosts interchanged and the second supercharge $\bar Q$ of the ${\cal N}=2$ superalgebra playing the role of the BRST charge in the dual picture.}
\begin{document}
\maketitle
%%%%%%%%%%%%%%%%%%%%%%%%%%%%%%%%%%%%%%%%%%%%%%%%%%%%%%%%%%%%%%%%%%%%%%%%%%%%%%%
\section{Introduction}

The remarkable confluence of ideas between theoretical physics and modern mathematics
has created an exciting symbiosis which has been extraordinarily productive on both sides, especially for quantum field theory, gravity and string theory on one hand, and differential geometry and algebraic topology on the other.

Continuing in this spirit, in \cite{grf} a new such connection has been proposed, which relates a topological version of nonrelativistic quantum gravity and the mathematical theory of the Ricci flow on Riemannian manifolds \cite{hrf,rfi,rf1,rf2,rf3,rf4,topping,muller}.    The construction of the appropriate version of quantum gravity was presented in \cite{grf} in stages.  First, a ``primitive'' topological gravity -- whose only dynamical field is the \textit{spatial} metric $g_{ij}$ -- was constructed, using the standard BRST techniques.  The existence of an extended $\CN=2$ BRST supersymmetry algebra was required, and there was no secondary gauge symmetry besides the topological deformations of $g_{ij}$.  The most convenient formulation of this theory is in terms of superfields on the appropriate $\CN=2$ superspace, and with global supercharges $Q$ and $\bar Q$.  In the next stage, the foliation-preserving spacetime diffeomorphisms were gauged, leading to new ingredients that bring the theory closer to the form that can accommodate the mathematical structure of Perelman's Ricci flow \cite{perel1,perel2,perel3}.

We chose such a two-stage construction of the theory in $\CN=2$ superspace because of its efficiency in identifying a suitable Lagrangian that serves our needs and makes contact with the Ricci flow.  However, this construction may also have left the reader somewhat mystified about several points.  First, the theory proposed in \cite{grf} has both topological gauge symmetries and spacetime gauge symmetries; yet, the two-stage construction in \cite{grf} uses a strange hybrid between a BRST gauge-fixed topological symmetry, and an unfixed gauge symmetry of foliation-preserving diffeomorphisms of spacetime.  It may be natural to seek a one-step construction, in which \textit{all} the gauge symmetries are handled uniformly, by BRST quantization.  Second of all, if the topological theory has a secondary gauge symmetry of spacetime diffeomorphisms, why were there no ghost-for-ghost fields required in \cite{grf}?  Such secondary ghost-for-ghosts are well-known to be essential in making sense of gauge theories with redundant gauge symmetries \cite{htbook}.  And finally, another puzzling feature emerges in the lapse sector:  The lapse superfields in \cite{grf} were introduced in the process of making the theory invariant under foliation-preserving time reparametrizations, and they were chosen to be nonprojectable superfields.  This nonprojectability was in turn important for making contact with Perelman's theory of the Ricci flow, since the role of the Perelman dilaton was found to be played by our nonprojectable lapse.  One might then question the number of local propagating degrees of freedom in such a theory:  The projectable time reparametrizations are not sufficient to remove the propagating polarization in the lapse, so is this theory even properly topological?  

The purpose of the present paper is to provide clarifying answers to all of these questions.  In particular, we show how the theory constructed in \cite{grf} can be consistently interpreted as a one-stage BRST gauge fixing of a topological theory whose original Lagrangian is zero.  In the process, we will be led to study in more detail the possible symmetries acting on the time dimension in the foliated spacetime, and on its $\CN=2$ supersymmetric extension to ``supertime.''  We will also clarify the notion of scaling dimensions in quantum gravity in situations when one cannot rely on a preferred, highly symmetric, background reference geometry.  Some of the lessons learned here may be of more general interest in relativistic and nonrelativistic quantum gravity, beyond the scope of the topological theory that we focus on in this paper.

This paper is organized as follows.  In the remainder of Section 1, we review the geometry (or, in the sense defined by J.L. Synge, ``chronometry'') of time in our topological quantum gravity from \cite{grf}, where the time dimension is a part of a $(1|2)$-dimensional supermanifold (the ``supertime''), review the structure of its superalgebra of symmetries, and summarize the structure of superfields in the theory.  In Section 2, we attempt to interpret the construction of our theory in \cite{grf} -- where it has been directly formulated in the global $\CN=2$ superspace -- as having been obtained from the BRST gauge-fixing procedure of a spacetime gauge symmetry including both topological deformations of fields and diffeomorphisms, in the component formalism.  To achieve this, we first review the structure of the BRST multiplets from \cite{grf}, and try to reconstruct the local gauge transformations it could have originated from.  This is done first for spacetime-dependent spatial diffeomorphisms in Section 2.1, and then for time reparametrizations in Section 2.2.  The surprising novelty here is that while the theory in \cite{grf} was constructed with spacetime foliation-preserving diffeomorphisms in mind, the gauge symmetry reconstructed from the BRST multiplets is actually larger, and contains arbitrary time reparametrizations!  This is discussed in Section 3, where we also show how this extended symmetry is consistent with the existence of a preferred foliation on spacetime: The action of spacetime diffeomorphisms on fields depends on a dimensionful parameter $c$ (which in the relativistic theory would be the speed of light), and our case corresponds to taking the ultralocal $c\to 0$ limit of this action of diffeomorphisms on fields.  The precise comparison between these gauge symmetries and the action of the BRST charge on fields requires an intriguing field-dependent twist, which we present in Section 3.3.  We then test this picture in Section 4, where we return to the superfield formalism of \cite{grf} and demonstrate that solving the superspace constraints precisely reproduces the structure of gauge symmetries suggested in Section 3.3.  In Section 4.2, we go to Wess-Zumino gauge, where all the higher components of the gauge superfields can be solved-for algebraically, and explain why our topological gravity of the Ricci flow does not lead to the ghost-for-ghost phenomenon:  In order to establish contact with Perelman's theory of the Ricci flow, our gauge symmetries are non-redundant.  While it would be in principle possible to consider theories with redundant symmetries and higher generation ghosts, it would take us away from the connection to Perelman's Ricci flow and is therefore outside of the scope of the present paper.  In Section 4.3 we point out that the same BRST theory has a dual interpretation, in terms of gauge-fixing a dual gauge symmetry, with the role of ghosts and antighosts reversed.  We do not, at present, understand the importance (if any) of this dual gauge symmetry.  We conclude in Section 5.

\subsection{Chronometry: The geometry of time}

At the center of our investigations in this paper will be the symmetries of time, and their interplay with the geometry and symmetries of the $D$-dimensional spatial slices of the $D+1$ dimensional spacetime.   Perhaps the more appropriate label for the topics studied here would be the ``chronometry of time and supertime''.  Indeed, already more than six decades ago \cite{synge}, one of the pioneers of the modern geometric approach to general relativity J.L. Synge advocated convincingly that measurements of space in general relativity are all secondary to the measurements of time, suggesting that spacetime ``geometry''%
\footnote{Geometry: From the Greek $\gamma\varepsilon\omega\mu\varepsilon\tau\!\rho\acute{\iota}\alpha$, ``measurement of earth or land''.}
should be more properly referred to as spacetime ``chronometry''%
\footnote{Chronometry: From the Greek $\chi\rho\acute{o}\nu o\varsigma$, ``time''; thus ``measurement of time''.}
(see also \cite{jammer}).  This suggestion appears particularly suitable in the context of nonrelativistic gravity \cite{mqc,lif}, where time indeed plays a privileged role, quite different from that of spatial dimensions.  Moreover, in the context of topological quantum gravity \cite{grf} associated with the nonrelativistic Ricci flow on Riemannian manifolds, time is far from being simply a real parameter $t$ taking values in $\R$.
In this nonrelativistic gravity, the ``geometry'' associated with time can be quite sophisticated, including a supersymmetrization of time -- but not of space -- to a supermanifold $\SM_0$ of dimension $(1|2)$, which we will naturally refer to as ``supertime.''  In this paper, we will find further geometric structure associated with time in nonrelativistic topological gravity, in particular a rather unusual form of local time reparametrization gauge symmetry that turns out to underlie the theory presented in \cite{grf}.  One can thus say that the present paper represents an investigation into this geometry of time, or ``chronometry,'' in topological quantum gravity of the Ricci flow.

\subsection{Supertime in topological quantum gravity of the Ricci flow}

We begin by reviewing some salient features of the nonrelativistic gravity theory presented in \cite{grf}.  The spacetime manifold $\CM$ is a $D+1$ dimensional manifold, on which we choose coordinates $t,x^i$, with $i=1,\ldots D$.  Even though we are primarily interested in the case of $3+1$ dimensions, it is not difficult to keep $D$ more general during our arguments.  We equip $\CM$ with the additional structure of a codimension-one foliation $\CM_\CF$ by spatial slices $\Sigma$ of constant time $t$.  Locally, this foliation defines a natural projection $\CM\to\CM_0$, where $\CM_0$ is the one-dimensional time, parametrized by $t$.  Typically, $\CM_0$ can be an open interval $I=(t_\textrm{init},t_\textrm{fin})$, and with the initial time $t_\textrm{init}$ and the final time $t_\textrm{fin}$ allowed to be finite or infinite; or it can be chosen to be a compact interval, when one considers either the initial-value problem, or transition amplitudes between two instants in time.

In the process of constructing a topological theory of gravity on $\CM$, one anticipates the existence of at least one supercharge $Q$, which will play the role of the BRST charge of the underlying gauge symmetry.  In \cite{grf}, we required in addition the existence of another supercharge $\bar Q$, the ``anti-BRST charge''.  They form a closed algebra with the generator of rigid time translations,
\be
Q^2=0,\qquad\bar Q^2=0,\qquad\{ Q,\bar Q\}=\p_t.
\label{eesusy}
\ee
This algebra of symmetries can be most efficiently implemented in the appropriate $\CN=2$ superspace.  Thus, the bosonic time manifold $\CM_0$ is promoted to an $\CN=2$ supersymmetric superspace $\SM_0$, the ``supertime'';  $\SM_0$ is parametrized by coordinates $(t,\theta,\bar\theta)$.  The supersymmetry algebra (\ref{eesusy}) is represented on $\CM_0$ by differential operators
\be
Q=\frac{\p}{\p\theta},\qquad\bar Q=\frac{\p}{\p\bar\theta}+\theta\frac{\p}{\p t}.
\ee
The spatial leaves $\Sigma$, on the other hand, play a much less prominent, almost a spectator role, and in particular they do not undergo any additional promotion to a supermanifold.  Thus, our full supersymmetrized spacetime $\SM$ is of dimension $(D+1|2)$, with a natural projection $\SM\to\SM_0$ to supertime, which in coordinates simply maps $(t,\theta,\bar\theta; x^i)\mapsto(t,\theta,\bar\theta)$.

The simplest topological quantum theory of gravity with (\ref{eesusy}) symmetry (referred to as the ``primitive'' theory in \cite{grf}) describes the dynamics of the spatial metric $g_{ij}(t,x^k)$.%
\footnote{Following the conventions chosen in \cite{grf}, we use Penrose's ``abstract index notation'' \cite{penrose} throughout for all our geometric objects on spacetime.  In this convention, the indices attached to geometric objects simply serve to specify which bundle the object is a section of, and they are not meant to indicate coordinate-dependent components.  Thus, in this sense, all our constructions are defined globally on spacetime, at least in the regions where no geometric singularity forms.}
In the superspace formulation, $g_{ij}$ is the lowest component of a superfield $G_{ij}$,
\be
G_{ij}=g_{ij}+\theta\psi_{ij}+\bar\theta\chi_{ij}+\theta\bar\theta B_{ij}.
\label{eegsup}
\ee
The component fields have a very natural interpretation:  $\psi_{ij}$ is the topological ghost, $\chi_{ij}$ the topological antighost, and $B_{ij}$ is its associated auxiliary field.  These are the standard BRST multiplets associated with the BRST gauge fixing of the topological gauge symmetry consisting of all local deformations of the metric,
\be
\delta g_{ij}=f_{ij}(t,x^k),
\ee
and with $Q$ playing the role of the BRST charge.  The antighost-auxiliary multiplet has been chosen such that the gauge-fixing condition that fixes the topological symmetry will also be given by a symmetric 2-tensor $\SF_{ij}$, which can depend on several coupling constants.  As we showed in \cite{grf}, for specific values of the couplings, this gauge fixing is equivalent to Hamilton's Ricci flow \cite{collrf,hrf}
\be
\SF_{ij}=\dot g_{ij}+2R_{ij}=0.
\ee
Thus, the primitive topological quantum gravity contains Hamilton's Ricci flow, but does not yet contain Perelman's generalization to the coupled flow of $g_{ij}$ and Perelman's ``dilaton'' field $\phi$.

In order to extend the theory to accommodate the Perelman-Ricci flow \cite{perel1,perel2,perel3}, in \cite{grf} we gauged the foliation-preserving diffeomorphism symmetry of spacetime.  In a bosonic theory, gauging spatial diffeomorphisms requires the introduction of a shift vector $n^i$, which plays the role of the gauge field for $\diff(\Sigma)$.  In order to make this gauging compatible with our $\CN=2$ supersymmetry, $n^i$ is also viewed as the lowest component of a superfield,%
\footnote{Note that we choose $N^i$ to be nonchiral; thus, we focus throughout this paper on the balanced, nonchiral theory referred to as ``Type B'' in \cite{grf}.}
\be
N^i=n^i+\theta\psi^i+\bar\theta\chi^i+\theta\bar\theta B^i.
\ee
The $N^i$ superfield covariantizes the time derivative with respect to the gauge symmetry $\diff (\Sigma)$ or, more precisely, its supersymmetric extension generated by the superfield $\Xi^i(t,x^j,\theta,\bar\theta)$ whose lowest component is $\xi^i(t,x^j)$.  For instance, the covariant derivative $\nabla_t$ on the metric superfield is given by
\be
\nabla_t G_{ij}\equiv\dot G_{ij}-\nabla_i N_j-\nabla_j N_i,
\ee
where $N_i\equiv G_{ij}N^j$, and $\nabla_i$ is the spatial covariant derivative that uses the Levi-Civit\`a connection built from $G_{jk}$.  The higher superpartners of $\xi^i$ in $\Xi^i$ can eventually be gauge-fixed by going to Wess-Zumino gauge \cite{grf}, reducing the gauge symmetries to the desired bosonic $\diff(\Sigma)$.  

In the next step, the $\CN=2$ supersymmetry requires that the odd derivatives $\RD$ and $\barRD$
\be
\RD=\frac{\p}{\p\theta}-\bar\theta\frac{\p}{\p t},\qquad\barRD =\frac{\p}{\p\bar\theta}
\ee
also be covariantized, which in turn required in \cite{grf} the introduction of gauge superfields $S^i$ and $\bar S^i$, 
\bea
S^i&=&\sigma^i+\theta \bar X^i+\bar\theta Y^i+\theta\bar\theta \rho^i,\\
\bar S^i&=&\bar\sigma^i+\theta X^i+\bar\theta\bar Y^i+\theta\bar\theta \bar\rho^i.
\eea
Since $S^i$ and $\bar S^i$ are Grassmann odd, the components $\sigma^i$, $\bar\sigma^i$, $\rho^i$ and $\bar\rho^i$ are fermions, and $X^i,\bar X^i, Y^i$ and $\bar Y^i$ are bosons.  This is a large and unwanted proliferation of component fields, and we expect them to be reduced by a number of covariant constraints.  Such constraints were identified in \cite{grf}:  
\bea
\label{eeconsts}
\RD S^i&=&S^k\p_k S^i,\\
\label{eeconstsbar}
\barRD\bar S^i&=&\bar S^k\p_k\bar S^i,\\
\label{eeconstshift}
N^i&=&-\barRD S^i-\RD\bar S^i+S^k\p_k\bar S^i+\bar S^k\p_k S^i.
\eea
With these shift-sector superfields $N^i$, $S^i$ and $\bar S^i$, one then has all the ingredients to covariantize the action of the primitive theory, to make it gauge invariant under (the $\CN=2$ superfield extension of) the spatial diffeomorphism gauge symmetry.

When we also gauge time reparametrizations, more superfields need to be introduced, leading to a rather complex set of constraints \cite{grf}.  We postpone discussing those until Section~\ref{sslapf} of this paper.

\section{Gauge symmetries in topological quantum gravity of the Ricci flow}

As we pointed out in the introduction, the superspace construction of the theory with gauged $\diff(\Sigma)$ in \cite{grf} leaves a few questions and loose ends.  It is a hybrid construction, in which the topological symmetry appears to have already been gauge fixed, but the secondary spatial diffeomorphism symmetry does not.  Thus, on one hand $Q$ is interpreted as the BRST charge for the topological symmetry, but from the perspective of $\diff(\Sigma)$ gauge symmetry, it appears to be treated as a supercharge of rigid supersymmetry.   It would be sensible to ask for a one-step interpretation of the resulting theory as a BRST gauge fixing of a clear list of gauge symmetries specified at the outset.  If such a one-step interpretation of our theory is possible, is the supercharge $Q$ the full BRST charge \textit{after} the gauging of spatial diffeomorphisms?  What are the underlying gauge symmetries that are being gauged?  Do they exhibit a redundancy, and if so, why did the procedure of \cite{grf} not require the presence of secondary ghost-for-ghosts?

To clarify these important questions, let us review how the standard process of constructing a topological field theory of the cohomological type, in the traditional component-field formulation \cite{ewcoho}, would work for nonrelativistic gravity.   

\subsection{Spatial diffeomorphisms and the shift vector $n^i$}
\label{ssdifni}

In the standard process of constructing a gauge theory, we first specify the fields, and then the gauge symmetries acting on those fields.  Then we apply the algorithm of BRST gauge fixing, which requires additional choices of gauge-fixing conditions.  In this section, we choose our fields to be the spatial metric $g_{ij}$ and the shift vector $n^i$.  We postpone the addition of the lapse function $n$ and time-reparametrization symmetries until Section~\ref{sslapf} below, and focus first on clarifying the gauging of the spatial diffeomorphisms $\diff(\Sigma)$.

In order to combine the topological gauge symmetries with the spatial diffeomorphisms, one could naturally start with a redundant system of symmetries,
\bea
\delta n^i&=&f^i(t,x^j)+\dot\xi^i+\xi^k\p_k n^i-\p_k\xi^i n^k,
\label{eenredu}
\\
\delta g_{ij}&=&f_{ij}(t,x^k)+\xi^k\p_k g_{ij}+\p_i\xi^k g_{kj}+\p_j\xi^k g_{ik}.
\label{eegredu}
\eea
Here $f_{ij}$ is the topological gauge symmetry acting by arbitrary local deformations of $g_{ij}$, $f^i$ generates a similar topological symmetry on the shift vector $n^i$, and $\xi^i$ is the standard generator of $\diff(\Sigma)$.  These symmetries are indeed redundant:  A given diffeomorphism transformation $\xi$ can be compensated by the choice of $f_{ij}$ and $f^i$ so that the combined transformation is zero.  In such cases, it is well-known that the proper BRST gauge fixing requires not only the fermionic ghosts $\Psi_{ij}$, $\Psi^i$ and $c^i$ associated with the gauge symmetries generated by $f_{ij}, f^i$ and $\xi^i$ (all of ghost number one), but also a bosonic ghost-for-ghost $\phi^i$ of ghost number two \cite{ewcoho,htbook}, which accounts for the redundancy of the original gauge symmetries (\ref{eenredu}) and (\ref{eegredu}).  These fields form a BRST multiplet of the BRST charge $\QB$ (with $\QB^2=0$), 
\bea
\QB\, n^i&=&\Psi^i+\dot c^i+c^k\p_k n^i-\p_k c^i n^k,\nonumber
\\
\QB\, g_{ij}&=&\Psi_{ij}+ c^k\p_k g_{ij}+\p_i c^k g_{kj}+\p_j c^k g_{ik},\nonumber\\
\QB\, c^i&=&c^k\p_k c^i+\phi^i,
\label{eeqbrst}\\
\QB\,\Psi^i&=&-\dot\phi^i+n^k\p_k\phi^i-\phi^k\p_k n^i+c^k\p_k\Psi^i+\Psi^k\p_kc^i,\nonumber\\
\QB\,\Psi_{ij}&=&-\phi^k\p_k g_{ij}-\p_i\phi^k g_{jk}-\p_j\phi^k g_{ik}+c^k\p_k\Psi_{ij}+\p_ic^k\Psi_{jk}+\p_jc^k\Psi_{ik},
\nonumber\\
\QB\,\phi^i&=&\phi^k\p_kc^i-c^k\p_k\phi^i.\nonumber
\eea
In contrast, in the hybrid two-step construction of \cite{grf}, no such ghost-for-ghost fields $\phi^i$ of ghost number two appeared to be necessary, and we wish to understand why.

To clarify this, we note first that this redundancy between topological and spacetime symmetries is almost inevitable in relativistic theories, especially in higher than the lowest few dimensions, if one wishes to maintain their relativistic symmetries.  In the case of our nonrelativistic gravity, however, there is a more elementary theory, which does not require ghost-for-ghosts.  The reasoning is reminiscent of the effects observed in the context of relativistic topological gravity in two dimensions in \cite{htcomm}.  Our non-redundant gauge symmetries will be
\bea
\delta n^i&=&\dot\xi^i+\xi^k\p_k n^i-\p_k\xi^i n^k,
\label{eencorr}
\\
\delta g_{ij}&=&f_{ij}+\xi^k\p_k g_{ij}+\p_i\xi^k g_{kj}+\p_j\xi^k g_{ik}.
\label{eegcorr}
\eea
We have simply left out from (\ref{eenredu}) the topological gauge symmetry generated by $f^i$ and acting on $n^i$.  Indeed, these symmetries are now non-redundant:  While the action of the spatial diffeomorphism $\xi^i$ on the spatial metric can be compensated for by a shift in the topological symmetry $f_{ij}$, no such compensation is possible in the action on $n^i$, and the two symmetries generated by $f_{ij}$ and $\xi^i$ are mutually independent.  Yet, clearly, these symmetries are powerful enough to eliminate all local propagating degrees of freedom, and the resulting theory will therefore still be ``topological'' in this sense.  The BRST charge acts as in (\ref{eeqbrst}), but with the fields $\Psi^i$ and $\phi^i$ omitted:
\bea
\QB\, n^i&=&\dot c^i+c^k\p_k n^i-\p_k c^i n^k,\nonumber\\
\QB\, g_{ij}&=&\Psi_{ij}+ c^k\p_k g_{ij}+\p_i c^k g_{kj}+\p_j c^k g_{ik},\nonumber\\
\QB\, c^i&=&c^k\p_k c^i,\\
\QB\,\Psi_{ij}&=&c^k\p_k\Psi_{ij}+\p_ic^k\Psi_{jk}+\p_jc^k\Psi_{ik}.\nonumber
\eea
Note one additional useful fact:  The symmetries are not only non-redundant, their Lie algebra decomposes into a direct sum of a symmetry acting solely on the shift $n^i$, and a topological symmetry acting only on $g_{ij}$.  This follows from a simple change of basis in the symmetry algebra, from $\xi^i$ and $f_{ij}$ to $\xi^i$ and $\hat f_{ij}$, with
\be
\hat f_{ij}\equiv f_{ij}+\xi^k\p_k g_{ij}+\p_i\xi^k g_{kj}+\p_j\xi^k g_{ik}.
\ee
This is a change of coordinates on the symmetry algebra, whose Jacobian is equal to one.  Calling by $\psi_{ij}$ the ghost associated with the shifted topological symmetry $\hat f_{ij}$, the BRST transformations now simplify to
\bea
\QB\, n^i&=&\dot c^i+c^k\p_k n^i-\p_k c^i n^k,\nonumber\\
\QB\, c^i&=&c^k\p_k c^i,\nonumber\\
\QB\, g_{ij}&=&\psi_{ij},
\label{eeqtbc}\\
\QB\,\psi_{ij}&=&0.\nonumber
\eea
The fact that such a shift exists, and that in the new basis the $\diff(\Sigma)$ symmetry acts trivially on $g_{ij}$, is important: The supermultiplets of $Q$ introduced in \cite{grf} in the process of gauging $\diff(\Sigma)$ exhibit the same type of decoupling between the two symmetries.  It is this form (\ref{eeqtbc}) of the BRST transformations that will be best suited for the comparison against our topological gravity constructed in Section~3 of \cite{grf}.  

We claim that the $\CN=2$ supersymmetry multiplets found in Section~3 of \cite{grf} are indeed equivalent to the standard multiplets associated with the BRST gauge fixing
of (\ref{eencorr}) and (\ref{eegcorr}), and that the supercharge $Q$ of \cite{grf} is simply the standard BRST charge of this one-step construction.  This will be best seen when we rewrite the theory developed in \cite{grf} in the component formalism.  In order to determine the independent component fields and their properties, we must first solve the constraints relating $N^i, S^i$ and $\bar S^i$.  Since the constraints (\ref{eeconsts}) and (\ref{eeconstsbar}) involve only $S^i$ and $\bar S^i$ respectively, they can be solved first, yielding
\bea
\label{eeconstss}
S^i&=&\sigma^i+\theta\sigma^k\p_k\sigma^i+\bar\theta Y^i+\theta\bar\theta(\dot\sigma^i+Y^k\p_k\sigma^i-\sigma^k\p_kY^i),\\
\label{eeconstsss}
\bar S^i&=&\bar\sigma^i+\theta X^i+\bar\theta\bar\sigma^k\p_k\bar\sigma^i+\theta\bar\theta (\bar\sigma^k\p_k X^i-X^k\p_k\bar\sigma^i).
\eea
Thus, the independent components of $S^i$ and $\bar S^i$ have been reduced to $\sigma^i, \bar\sigma^i, X^i$ and $Y^i$, transforming as follows under $Q$:
\bea
Q\, \sigma^i&=&\sigma^k\p_k\sigma^i,\\
Q\, Y^i&=&-\dot\sigma^i-Y^k\p_k\sigma^i+\sigma^k\p_kY^i,\\
Q\,\bar\sigma^i&=&X^i,\\
Q\,X^i&=&0.
\eea

These are not yet the appropriate component fields of our desired supersymmetry multiplets, as we still need to solve the constraint (\ref{eeconstshift}) that expresses the components of $N^i$ in terms of the independent components in $S^i$ and $\bar S^i$.  First, solving (\ref{eeconstshift}) at the lowest component gives
\be
\label{eecompshift}
n^i=-Y^i-X^i+\sigma^k\p_k\bar\sigma^i+\bar\sigma^k\p_k\sigma^i.
\ee

Note that because $X^i$ is $Q$ invariant, it is our candidate for the auxiliary field in a trivial BRST multiplet, paired up with $\bar\sigma^i$ which will play the role of the antighost.  This means that the correct way of reading (\ref{eecompshift}) is to interpret $Y^i$ as a composite field, determined by this equation in terms of the four independent component fields $n^i, X^i, \sigma^i$ and $\bar\sigma^i$,
\be
Y^i=-n^i-X^i+\sigma^k\p_k\bar\sigma^i+\bar\sigma^k\p_k\sigma^i.
\label{eeysol}
\ee
This is the sense in which $Y^i$ should be understood as a composite field when it appears in expressions such as (\ref{eeconstss}).  

The remaining components of $N^i$ can now be determined in terms of the independent fields $n^i, X^i, \sigma^i$ and $\bar\sigma^i$ by evaluating the higher-order terms in (\ref{eecompshift}).  They are found to be
\bea
\label{eecompshpsi}
\psi^i&=&\dot\sigma^i+\sigma^k\p_k n^i-\p_k\sigma^i n^k,\\
\label{eecompshchi}
\chi^i&=&\dot{\bar\sigma}^i+\bar\sigma^k\p_k n^i-\p_k\bar\sigma^i n^k,\\
B^i&=&-\dot X^i+n^k\p_k X^i-X^k\p_kn^i+\dot\sigma^k\p_k\bar\sigma^i+\sigma^j\p_jn^k\p_k\bar\sigma^i-n^j\p_j\sigma^k\p_k\bar\sigma^i\nonumber\\
&&\ {}+\bar\sigma^k\p_k\dot\sigma^i+\bar\sigma^k\p_k\sigma^j\p_jn^i+\bar\sigma^k\sigma^j\p_k\p_jn^i-\bar\sigma^k\p_kn^j\p_j\sigma^i-\bar\sigma^kn^j\p_k\p_j\sigma^i.
\eea
Thus, the entire sector of superfields $N^i$, $S^i$ and $\bar S^i$ reduces in components to four independent fields, belonging to two BRST multiplets of the BRST charge $Q$, with BRST transformations
\bea
Q\, n^i&=&\dot\sigma^i+\sigma^k\p_kn^i-n^k\p_k\sigma^i,
\label{eeqone}\\
Q\, \sigma^i&=&\sigma^k\p_k\sigma^i,
\label{eeqtwo}
\eea
and
\bea
Q\,\bar\sigma^i&=&X^i,
\label{eeqthree}\\
Q\,X^i&=&0.
\label{eeqfour}
\eea
Clearly, $\sigma^i$ is the ghost of the spatial diffeomorphism symmetry, and should be identified with $c^i$ in (\ref{eeqbrst}), while $\bar\sigma^i$ and $X^i$ form the trivial BRST multiplet consisting of an antighost and an auxiliary field.  In addition to these two BRST multiplets, the theory that was constructed in Section~3 of \cite{grf} by gauging spatial diffeomorphisms of the primitive topological gravity contains also the two BRST multiplets in the metric superfield of (\ref{eegsup}),
\bea
Q\,g_{ij}=\psi_{ij},\qquad Q\,\psi_{ij}=0,
\label{eeqfive}\\
Q\,\chi_{ij}=-B_{ij},\qquad Q\,B_{ij}=0.
\label{eeqsix}
\eea
We see that these component supermultiplets (\ref{eeqone}), (\ref{eeqtwo}) and (\ref{eeqfive}) of $Q$ match exactly the multiplets (\ref{eeqtbc}) of $\QB$, which we obtained by the one-step BRST gauge fixing of the non-redundant gauge symmetries (\ref{eencorr}) and (\ref{eegcorr}).  And the remaining multiplets of $Q$, listed in (\ref{eeqthree}), (\ref{eeqfour}) and (\ref{eeqsix}), are the standard antighost-auxiliary BRST multiplets ready for the implementation of our gauge fixing choice.  In this way, the two-step construction in Section~3 of \cite{grf} can indeed be consistently interpreted as the standard one-step gauge fixing of a gauge theory with non-redundant gauge symmetries, with $Q$ identified as the BRST charge.

\subsection{Time reparametrizations and the lapse function $n$}
\label{sslapse}

In Section~4 of \cite{grf}, the theory was further extended by gauging foliation-preserving time reparametrizations, and adding the appropriate $\CN=2$ supersymmetrization of the lapse function $n$.  In order to clarify whether this extended theory can also be interpreted as a one-step BRST gauge fixing of appropriate non-redundant gauge symmetries, we need to overcome additional challenges which were absent in the theory with $\diff(\Sigma)$ gauge symmetry discussed above.  We begin by briefly reviewing the symmetries and fields introduced in \cite{grf}.  

The gauge symmetries of spacetime-dependent spatial diffeomorphisms $\diff(\Sigma)$ are extended to all foliation-preserving diffeomorphisms of the $D+1$ dimensional spacetime $\CM_\CF$.  Besides the generators $\xi^i(t,x^k)$ that were considered in the previous paragraph, this extended symmetry also contains time-dependent time reparametrizations, generated by $f(t)$:
\bea
\delta n&=&\dot f n+f\dot n,\nonumber\\
\delta n^i&=&\dot f n^i+f\dot n^i,\\
\delta g_{ij}&=&f\dot g_{ij}.\nonumber
\eea
Here we have introduced besides the spatial metric $g_{ij}$ and the shift vector $n^i$ also the lapse function $n$, which plays the role of the gauge field associated with time reparametrizations:  More precisely, it is $\log n$ that transforms as a gauge field: $\delta\, \log n=\dot f+\ldots$.  While it is mathematically consistent in nonrelativistic gravity to declare the field $n$ to be a projectable field $n(t)$, it is physically more interesting to allow the lapse to be a nonprojectable field $n(t,x^i)$, \ie , to promoting it to a spacetime-dependent field.  Allowing the lapse to be nonprojectable was indeed crucial for us in \cite{grf} in establishing contact with Perelman's theory of the Ricci flow, since we found that the role of Perelman's ``dilaton'' field is played by the nonprojectable lapse field.  

In Section~4 of \cite{grf}, we constructed a nonrelativistic gravity theory which is gauge invariant under $f(t)$ and consistent with our requirement of rigid $\CN=2$ supersymmetry of supertime, using the standard techniques for gauging symmetries in superspace.  We promoted $n(t,x^i)$ to a nonchirial, nonprojectable superfield, whose lowest component is the nonprojectable lapse, $N=n+\ldots$ \cite{grf}.  In fact, in \cite{grf} we found it convenient to introduce the \textit{inverse} lapse superfield, $E=1/N$, whose lowest component is the nonprojectable field $e\equiv 1/n$,
\be
E=e+\theta\psi+\bar\theta\chi+\theta\bar\theta B.
\ee
This $E$ superfield then covariantizes the time derivative under (the supersymmetric extensions of) the $f(t)$ symmetries, in the way explained in detail in \cite{grf}.

Since the odd derivatives also need to be covariantized, two more superfields $\Theta$ and $\bar\Theta$ had to be introduced in addition to $E$; they are also nonprojectable, nonchiral, but odd, and we introduce the following notation for their components:%
\footnote{The minus signs introduced here in the process of naming the component fields will simplify some future component formulas below, and they also make the parallel between the odd superfields $\Theta, \bar\Theta$ of the lapse sector and the odd superfields $S^i, \bar S^i$ of the shift sector more explicit.}
\bea
\Theta&=&-\nu-\theta\bar w-\bar\theta z-\theta\bar\theta\rho,\\
\bar\Theta&=&-\bar\nu-\theta w-\bar\theta \bar z-\theta\bar\theta\bar\rho.
\eea
As in the case of spatial diffeomorphisms discussed in the previous section of the present paper, this proliferation of component fields is reduced to the minimal independent set by suitable constraints, which now involve both the shift sector and the lapse sector superfields \cite{grf}:
\bea
\RD\Theta-S^k\p_k\Theta&=&-\Theta(\dot\Theta-N^k\p_k\Theta),
\label{eelcon1}\\
\barRD\bar\Theta-\bar S^k\p_k\bar\Theta&=&-\bar\Theta(\dot{\bar\Theta}-N^k\p_k\bar\Theta),
\label{eelcon2}
\eea
and
\be
E=1-\RD\bar\Theta+S^k\p_k\bar\Theta-\barRD\Theta+\bar S^k\p_k\Theta-\Theta(\dot{\bar\Theta}-N^k\p_k\bar\Theta)-\bar\Theta(\dot\Theta-N^k\p_k\Theta).
\label{eelcon3}
\ee
In addition to these mixed constraints, the shift sector superfields $S^i, \bar S^i$ and $N^i$ still satisfy constraints (\ref{eeconsts}), (\ref{eeconstsbar}) and (\ref{eeconstshift}).  These constraints appear somewhat more involved than in the case of the shift sector alone, and we postpone solving them until Section~\ref{ssconstr}.

Once the nonprojectable lapse sector has been introduced, the pattern that we uncovered in the case with $\diff(\Sigma)$ gauge symmetry will continue to be valid if the BRST multiplets in the lapse sector can be interpreted as having originated in BRST gauge fixing of a hidden symmetry associated with time.  Note that since the lapse sector contains only nonprojectable fields, this hidden symmetry cannot be simply the projectable time reparametrizations generated by $f(t)$.  This hidden symmetry will have to be nonprojectable, generated by some space and time dependent parameter $\zeta(t,x^i)$.   Is there such a hidden symmetry, and if so, what is its geometric interpretation?

The first guess might be that $\zeta(t,x^i)$ should perhaps be some kind of nonprojectable time reparametrization symmetry, which would extend the symmetry of spatial diffeomorphisms generated by $\xi^i(t,x^k)$.  This symmetry would have to satisfy several stringent requirements.  For example, it would have to act on $n^i$ (and also on $g_{ij}$) trivially, or more accurately, in a way which can be absorbed into a redefinition of the spatial diffeomorphisms $\xi^i$ (and the topological gauge transformations $\hat f_{ij}$).  This phenomenon is analogous to what happened above, when we found the basis of symmetry generators in which $\xi^i$ acts trivially on $g_{ij}$, and therefore $\QB\,g_{ij}$ is independent of $c^i$ (see (\ref{eeqtbc})).  Is it possible to construct such a symmetry $\zeta$?  We could try to read off the rules directly from the transformation properties of the component fields under $Q$, assuming that they will conform to the interpretation as having come from the BRST fixing of a symmetry, with $Q$ being the BRST charge.  However, we find it more instructive to look first for available symmetries in the context of previously studied symmetries of spacetime, which we will do in the next section.  In the process, we will learn a few surprising facts which might be of more general interest for quantum gravity, beyond the applications to the main subject of this paper.  The reader who is solely interested in our topological nonrelativistic gravity construction can proceed directly to Section~\ref{sslapf}.

\section{Spacetime diffeomorphisms in relativistic and nonrelativistic gravity}
\label{ssdiffeos}

Consider the spacetime diffeomorphism generated by $\xi^\mu$, which we decompose into $\zeta(t,x^i)\equiv\xi^0$ and $\xi^i(t,x^k)$.  On the ADM decomposition \cite{adm}
of the relativistic spacetime metric into the spatial metric plus the lapse and shift,%
\footnote{For some background on the ADM decomposition, see, \eg , Chapter 21.4 of \cite{mtw}, Chapter 2.2 of \cite{carlip}, or \cite{mqc,lif}.}
the relativistic spacetime diffeomorphisms act via
\bea
\label{eestdiffn}
\delta n&=&\xi^k\p_k n+\zeta\dot n+(\dot\zeta -n^k\p_k\zeta)n,\\
\label{eestdiffni}
\delta n^i&=&\dot\xi^i+\xi^k\p_k n^i-\p_k\xi^in^k+\zeta\dot n^i+\dot\zeta n^i-\p_k\zeta n^kn^i
-\p_k\zeta g^{ik}n^2,\\
\label{eestdiffg}
\delta g_{ij}&=&\xi^k\p_k g_{ij}+\p_i\xi^k g_{kj}+\p_j\xi^k g_{ik}+\zeta\dot g_{ij}+(\p_i\zeta g_{jk}+\p_j\zeta g_{ik})n^k.
\eea
Can the standard action of the time diffeomorphism $\zeta(t,x^i)$ on $n^i$ be absorbed into the spatial diffeomorphism $\xi^i$?  Almost:  If the last term in (\ref{eestdiffni}), proportional to $g^{ij}n^2$, weren't there, it would be possible to absorb the action of $\zeta$ into that of the spatial diffeomorphism by changing the basis in the Lie algebra by first shifting $\xi^i=\hat\xi^i-\zeta n^i$ and then using $\hat\xi^i$ and $\hat\zeta\equiv\zeta$ as the new basis in the space of symmetry generators.  Note that this is again a change of variables whose Jacobian is equal to one.  The resulting algebra would then satisfy the three stringent requirements necessary for the possible matching with the superspace construction of our nonrelativistic gravity from Section~4 of \cite{grf}:
\begin{itemize}
\item
The shifted time reparametrization generators $\hat\zeta$ should act trivially on $n^i$;
\item
The shifted spatial diffeomorphism generators $\hat\xi^i$ should form the gauge algebra of $\diff(\Sigma)$;
\item
The commutator of a shifted spatial diffeomorphism $\hat\xi^i$ and a shifted time reparametrization $\hat\zeta$ should yield another shifted time reparametrization, and therefore act trivially on $n^i$.
\end{itemize}
In the standard spacetime diffeomorphism symmetry algebra, the last term on the right-hand side of (\ref{eestdiffni}) represents an obstruction to these three requirements.  In order to clarify the status of this obstruction term, and to see under what circumstances it can be set to zero, we will revisit the nonrelativistic decomposition of the spacetime diffeomorphism symmetries, and their nonrelativistic limits.

\subsection{Scales and scaling dimensions}
\label{ssscales}

In order to discuss the nonrelativistic decomposition of a relativistic spacetime, we wish to re-introduce the measurement of time and space in unrelated units: $L$ and $T$.  First, we recall a few elementary facts about scaling and dimensions in quantum field theory, and propose a simple refinement to the case when the system contains a dynamical spacetime metric.  

In quantum field theory in a Minkowski spacetime, scaling dimensions of various objects (quantum fields, spacetime derivatives, composite operators built out of them) play a central role, controlled by the concept of the renormalization group (RG).  In relativistic field theories, it is natural to set the speed of light $c=1$, and all dimensions are then expressed in terms of one dimensionful scale.  By convention, largely for historical reasons, in particle physics this scale is typically selected to be the momentum scale, or equivalently the inverse length scale when one sets $\hbar =1$.  In momentum units, Cartesian spacetime coordinates $x^\mu$ of the Minkowski spacetime are of dimension $-1$, and the derivatives $\p_\mu$ are of dimension $1$.  Dimensions of quantum fields and various composite operators then depend on the type of theory in question, typically controlled by an RG fixed point.  Sometimes, however, there is a strong geometric reason to assign a given field a particular scaling dimension.  For example, in general relativity, the components $g_{ij}$ of the spacetime metric relate the proper length element $ds$ to the coordinate length elements $dx^i$ via $ds^2=g_{ij}dx^idx^j$.  It is only sensible to consider $ds^2$ to be of dimension $-2$ in momentum units, hence implying the often-quoted fact that for the purposes of power counting, the components $g_{ij}$ of the metric are dimensionless.  This picture is of course well-known, and almost universally accepted, but as we explain below, in need of a slight conceptual modification in theories with dynamical inhomogeneous geometries.  

Nonrelativistic gravity of the Lifshitz type \cite{mqc,lif}, and its topological counterpart studied here, belong to the class of quantum field theories with two distinct, \textit{a priori} unrelated scales:  a length scale $\LS$, and a time scale $\TS$, or their inverses: an energy scale $\ES=1/\TS$ and a momentum scale $\MS=1/\LS$ (having again set $\hbar=1$).  The relativistic strategy of assigning dimensions needs to be refined:  We can declare that the time coordinate carries dimension $\TS\equiv\ES^{-1}$, and the spatial coordinates $x^i$ are of dimension $\LS\equiv\MS^{-1}$.  (This was also the convention we used for assigning classical scaling dimensions to various objects in topological gravity of the Ricci flow in Section~2 of \cite{grf}.)  In addition, one simplifying convention seems appropriate:  Instead of writing the dimension $[\CO]$ of an object $\CO$ multiplicatively in terms of powers of $\TS$ and $\LS$ (or $\ES$ and $\MS$), we will simply write $[\CO]=(n,m)$ when $\CO$ is of dimension $\ES^n\MS^m$ in energy and momentum units.  With this convention we have, for instance,
\be
   [\p/\p t]=(1,0),\qquad [\p_i]=(0,1).
   \label{eeoldder}
\ee
Applying this analysis to the fields of nonrelativistic gravity, we thus obtain the ``standard'' dimension assignments:
\be
[g_{ij}]=(0,0),\qquad [n]=(0,0),\qquad [n^i]=(1,-1).
\label{eeoldf}
\ee
They simply follow from the fact that in nonrelativistic gravity, the geometry of spacetime is characterized by an invariant line element $d\sigma$ of spatial distance, defined via
\be
d\sigma^2=g_{ij}(dx^i+n^idt)(dx^j+n^jdt),
\ee
which should naturally be of dimension $[d\sigma]=(0,-1)$, and an independent invariant element of time,
\be
d\tau=n\,dt,
\ee
naturally of dimension $(-1,0)$.

In theories with two (or more) \textit{a priori} unrelated scales, the ``one-scale'' physics of the renormalization group scaling emerges in the vicinity of an RG fixed point, where the two scales become locked together by a scaling relation
\be
\label{eedynce}
E=M^z,
\ee
involving the so-called dynamical critical exponent $z$.  This exponent is characteristic of the specific RG fixed point, not of the theory itself -- it is indeed possible to have a theory with more than one RG fixed points, with distinct values of $z$.  In the vicinity of a given fixed point, one is then free to use (\ref{eedynce}) to express the RG scaling properties of the system in terms of just one independent scale (say $E$).

In the present context of topological gravity of the Ricci flow, and more generally in nonrelativistic gravity with a dynamical metric, it has become increasingly apparent that a slightly modified strategy for assigning dimensions would be more logical. There are several reasons for that.  In the flat Minkowski spacetime, or in a theory with a preferred and highly symmetric background, it may be natural to continue with the picture developed in quantum field theory, and see the spacetime coordinates $x^\mu$ or the derivatives $\p_\mu$ as the carriers of a scaling dimension:  After all, the idea of scaling is often phrased as a study of the properties of the system under the rescalings $x^\mu\mapsto bx^\mu$ for constant $b$.  However, this intuition is increasingly problematic in a theory whose background geometries are often highly inhomogeneous, such as our topological theory of the Ricci flow, whose path integral localizes to the solutions of generalized Ricci flow equations with arbitrarily inhomogeneous initial conditions.  On such backgrounds, there is no analog of a ``preferred coordinate system'' $x^\mu=(t,x^i)$.  Instead, we prefer to treat all coordinate systems equally; this will make sense only if we assign the scaling dimension $(0,0)$ to both $t,x^i$ and the derivatives $\p_t,\p_i$.  Only then the elementary covariance of our rules will be restored, and the concept of scaling dimensions will not have to rely on the existence of a preferred system of coordinates associated with a maximally symmetric ground-state geometry.

In this modified picture, the dimensions are now carried by the physical spatial length element $d\sigma^2$ and time element $d\tau$,
\be
[d\sigma^2]'=(0,-2),\qquad [d\tau]'=(-1,0),
\ee
just as we declared above.  However, since now $dx^i$ and $dt$ are declared dimensionless, this implies that the old-fashioned rules (\ref{eeoldder}) and (\ref{eeoldf}) are modified to
\be
[g_{ij}]'=(0,-2),\qquad [n]'=(-1,0),\qquad [n^i]'=(0,0),
\ee
and
\be
[\p/\p t]'=(0,0),\qquad [\p_i]'=(0,0).
\ee
We have denoted the dimension in the modified counting system by $[\ ]'$, to distinguish it from the dimension $[\ ]$ in the old-fashioned system.  

One can see that both systems of assigning dimensions are for all physical purposes equivalent to each other, for instance they lead to the same dimension counting rules for the terms that can appear in the action.  However, the modified system in which the underlying spacetime coordinates are naturally dimensionless is mathematically more accurate and conceptually cleaner that the old-fashioned one, since it is manifestly consistent with nonlinear changes of coordinates, and does not lead to apparent violations of the additivity properties of the classical scaling dimensions during such nonlinear coordinate transformations.

After this detour, we can now return to the study of the relativistic diffeomorphism symmetry in the ADM decomposition, in a theory with two scales $L$ and $T$.  Restoring the dimensions of various objects in the symmetry algebra (\ref{eestdiffn}), (\ref{eestdiffni}) and (\ref{eestdiffg}), we see that the last term on the right-hand side of (\ref{eestdiffni}) is the only one whose dimension does not match the rest of the terms, and therefore a dimensionful constant is needed to provide the conversion.  This constant is of course the speed of light $c$, of dimension $[c]=[c]'=(1,-1)$, which was set equal to one in the relativistic theory.  Restoring $c$, the algebra is now
\bea
\delta n&=&\xi^k\p_k n+\zeta\dot n+(\dot\zeta-n^k\p_k\zeta) n,\nonumber\\
\delta n^i&=&\dot\xi^i+\xi^k\p_k n^i-\p_k\xi^in^k+\zeta\dot n^i+\dot\zeta n^i-\p_k\zeta n^kn^i
-c^2\p_k\zeta g^{ik}n^2,
\label{eecalg}\\
\delta g_{ij}&=&\xi^k\p_k g_{ij}+\p_i\xi^k g_{kj}+\p_j\xi^k g_{ik}+\zeta\dot g_{ij}+(\p_i\zeta g_{jk}+\p_j\zeta g_{ik})n^k.\nonumber
\eea
It is an elementary but valuable exercise to verify why this insertion of $c^2$ makes the dimensions right in both of our dimension-counting systems.

Now we can return to our quest for a symmetry that would match the ingredients required by the topological nonrelativistic gravity of \cite{grf}.  As we saw at the beginning of Section~2 above, it is precisely this $c^2$ term that represents an obstruction against what we need.  This term can be simply eliminated by taking the $c\to 0$ limit, well-known in the literature \cite{tultra,hultra} (see also \cite{gen} for the nonrelativistic context) as the ``ultralocal'' limit of the relativistic diffeomorphism symmetries of general relativity.

\subsection{The $c\to 0$ limit}

Taking $c\to 0$ in (\ref{eecalg}) will allow us to follow the strategy outlined at the beginning of Section~\ref{ssdiffeos}, and to make the time transformation generated by $\zeta$ act trivially on the shift vector $n^i$.  Before taking these steps, let us take a closer look at this ``ultralocal'' limit of $c\to 0$, especially in the light of our improved prescription for assigning scaling dimensions.  According to this prescription, our coordinates $x^i,t$ are dimensionless, and therefore the generators $\xi^i,\zeta$ of the corresponding spacetime symmetries are dimensionless as well.  Yet, the algebra defined by the transformation rules (\ref{eecalg}) clearly depends on $c$.  Indeed, we are planning to take the contraction of the symmetries by sending $c\to 0$.  One might therefore think that the structure constants of this symmetry algebra, which one could obtain from the commutation relations of the transformations in (\ref{eecalg}), will depend on $c$.  However, that clearly represents a puzzle:  If the generators $\xi^i$ and $\zeta$, and all the derivatives $\p_t$ and $\p_i$ that can appear in the commutation relations are all dimensionless, how could the commutator of two such transformations possibly depend on a dimensionful constant such as $c$?

The resolution is simple, yet perhaps a little surprising.  A direct calculation reveals that regardless of the value of $c$, the commutation relations of the transformations in (\ref{eecalg}) are independent of $c$:  The commutator of a transformation $\delta_1$ generated by $\xi^i_1$ and $\zeta_1$ with the transformation generated by $\xi^i_2$ and $\zeta_2$,
\be
[\delta_1,\delta_2]=\delta_3,
\ee
is the transformation $\delta_3$ generated by the following $\xi^i_3$ and $\zeta_3$,
\bea
\zeta_3&=&\zeta_1\dot\zeta_2-\zeta_2\dot\zeta_1+\xi^k_1\p_k\zeta_2-\xi^k_2\p_k\zeta_1,\\
\xi^i_3&=&\xi^k_1\p_k\xi^i_2-\xi^k_1\p_k\xi^i_2+\zeta_1\dot\xi^i_2-\zeta_2\dot\xi^i_1.
\eea
These are indeed the commutation relations of the general spacetime diffeomorphisms, familiar from the relativistic theory of gravity.

Consider now the theory of bosonic gravity which would be invariant under the $c\to 0$ limit of the symmetries specified in (\ref{eecalg}).  This is of course the theory known in the literature as the ``ultralocal'' theory of gravity \cite{tultra,hultra}, proposed originally as a possible strong-coupling limit of general relativity.  Do the relativistic commutation relations of the symmetry generators mean that this theory is somehow relativistic?  Not in the standard sense of full relativistic general covariance:  While the commutation relations appear relativistic, the realization of the symmetries on the ADM decomposition of the metric does depend on $c$.  Consequently, Lagrangians that are invariant under this realization of the symmetries will be contractions of the standard Lagrangians of general relativity.  In particular, the lowest-derivative Lagrangian invariant under these symmetries,
\be
S=\frac{1}{\kappa^2}\int dt\,d^Dx\left\{\frac{\sqrt{g}}{n}\left(g^{ik}g^{j\ell}-g^{ij}g^{k\ell}\right)\left(\dot g_{ij}-\nabla_in_j-\nabla_jn_i\right)\left(\dot g_{k\ell}-\nabla_kn_\ell-\nabla_\ell n_k\right)-2n\sqrt{g}\Lambda\right\},
\label{eeaultra}
\ee
contains the standard kinetic term for the spatial metric with two time derivatives, but no spatial-derivative term consistent with it.  This of course is the standard result known from \cite{tultra,hultra}.  It is intriguing, however, that the dependence of this theory on the preferred foliation by constant time slices appears here at the level of the dynamical metric fields, and not at the level of the underlying symmetries of the differentiable structure of spacetime with dimensionless generators $\xi^i$ and $\zeta$.  

\subsection{Decoupling time reparametrizations from the shift vector}

We can now finally propose a meaningful candidate for the spacetime gauge symmetry that underlies the construction of the topological nonrelativistic theory in Section~4 of \cite{grf}.  

After performing the change of basis $\hat\xi^i=\xi^i+\zeta n^i$ and $\hat\zeta=\zeta$ in the generators of the $c=0$ transformation rules, which is again a change of basis whose Jacobian is one, we get
\bea
\hat\delta n&=&\hat\xi^k\p_k n+\hat\zeta(\dot n-n^k\p_k n)+(\p_t\hat\zeta-n^k\p_k\hat\zeta)n,\\
\hat\delta n^i&=&\p_t\hat\xi^i+\hat\xi^k\p_k n^i-\p_k\hat\xi^in^k,\\
\hat\delta g_{ij}&=&\hat\xi^k\p_k g_{ij}+\p_i\hat\xi^k g_{kj}+\p_j\hat\xi^k g_{ik}+\hat\zeta(\dot g_{ij}-n^k\p_k g_{ij}-\p_i n^kg_{kj}-\p_j n^kg_{ik}).
\eea
These transformations satisfy the following commutation relations:  The commutator of a transformation $\hat\delta_1$ generated by $\hat\xi^i_1$ and $\hat\zeta_1$ with a transformation $\hat\delta_2$ generated by $\hat\xi^i_2$ and $\hat\zeta_2$,
\be
[\hat\delta_1,\hat\delta_2]=\hat\delta_3,
\ee
is the transformation $\hat\delta_3$ generated by 
\bea
\hat\zeta_3&=&\hat\zeta_1\,(\p_t\hat\zeta_2-n^k\p_k\hat\zeta_2)-\hat\zeta_2\,(\p_t\hat\zeta_1-n^k\p_k\hat\zeta_1)+\hat\xi^k_1\p_k\hat\zeta_2-\hat\xi^k_2\p_k\hat\zeta_1,\\
\hat\xi^i_3&=&\hat\xi^k_1\,\p_k\hat\xi^i_2-\hat\xi^k_1\,\p_k\hat\xi^i_2.
\eea
We will simply refer to this symmetry algebra as $\CG$.  Note that the structure ``constants'' of $\CG$ in this basis are field-dependent, due to the explicit appearance of $n^k$.  This field dependence of the structure constants could of course be undone by ``unshifting'' the basis and representing the transformations using the original generators $\xi^i$ and $\zeta$.  However, in the context of the BRST gauge-fixing in our topological gravity from \cite{grf}, it is the shifted representation in terms of $\hat\xi^i$ and $\hat\zeta$ that will make a natural appearance.  Note also that the action of the algebra $\CG$ on $n$ and $n^i$ satisfies all the necessary requirements listed in our three bullet points at the beginning of Section~\ref{ssdiffeos}, needed to be a candidate for our underlying spacetime symmetry for the topological gravity of \cite{grf}:  The subalgebra in $\CG$ generated by $\hat\xi^i$ is the standard Lie algebra of spatial diffeomorphisms, while the subalgebra generated by $\hat\zeta$ is an ideal in $\CG$, and this ideal acts trivially on the shift vector $n^i$.

Besides the symmetries generated by $\hat\zeta$ and $\hat\xi^i$, we also include the original topological symmetries $f_{ij}$ acting on $g_{ij}$, as we did in (\ref{eegredu}).  Putting all the symmetries together, we reach the following algebra: 
\bea
\hat\delta n&=&\hat\xi^k\p_k n+\hat\zeta(\dot n-n^k\p_k n)+(\p_t\hat\zeta-n^k\p_k\hat\zeta)n,\\
\hat\delta n^i&=&\p_t\hat\xi^i+\hat\xi^k\p_k n^i-\p_k\hat\xi^in^k,\\
\hat\delta g_{ij}&=&f_{ij}+\hat\xi^k\p_k g_{ij}+\p_i\hat\xi^k g_{kj}+\p_j\hat\xi^k g_{ik}+\hat\zeta(\dot g_{ij}-n^k\p_k g_{ij}-\p_i n^kg_{kj}-\p_j n^kg_{ik}).
\eea
Finally, redefining the topological transformation to
\be
\hat f_{ij}\equiv f_{ij}+\hat\xi^k\p_k g_{ij}+\p_i\hat\xi^k g_{kj}+\p_j\hat\xi^k g_{ik}+\hat\zeta(\dot g_{ij}-n^k\p_k g_{ij}-\p_i n^kg_{kj}-\p_j n^kg_{ik}),
\ee
which is again a change of basis whose Jacobian is equal to one, we bring the symmetries to the following simple form:
\bea
\hat\delta n&=&\hat\xi^k\p_k n+\hat\zeta(\dot n-n^k\p_k n)+(\p_t\hat\zeta-n^k\p_k\hat\zeta)n,\nonumber\\
\hat\delta n^i&=&\p_t\hat\xi^i+\hat\xi^k\p_k n^i-\p_k\hat\xi^in^k,
\label{eefinalsym}\\
\hat\delta g_{ij}&=&\hat f_{ij}.\nonumber
\eea
Note that the symmetries generated by $\hat\zeta$, $\hat\xi^i$ and $\hat f_{ij}$ are non-redundant.  When we treat these gauge symmetries using the BRST formalism, only first-generation ghosts will be needed, and no ghost-for-ghosts.  Introducing the ghosts $c$, $c^i$ and $\psi_{ij}$ associated with our symmetry generators $\hat\zeta$, $\hat\xi^i$ and $\hat f_{ij}$, the BRST transformations of this BRST multiplet are
\bea
\QB\, n^i&=&\dot c^i+c^k\p_k n^i-\p_k c^i n^k,\nonumber\\
\QB\, c^i&=&c^k\p_k c^i,\nonumber\\
\QB\, n&=&c(\dot n-n^k\p_k n)+(\dot c-n^k\p_kc)\,n+c^k\p_kn,\nonumber\\
\QB\, c&=&c(\dot c-n^k\p_kc)+c^k\p_kc,\label{eeqtbc2}\\
\QB\, g_{ij}&=&\psi_{ij},\nonumber\\
\QB\,\psi_{ij}&=&0.\nonumber
\eea
It is this BRST algebra that we now wish to compare to the structure of the $Q$ supermultiplets of the topological nonrelativistic gravity from Section~4 of \cite{grf}.

\section{Superfields in topological quantum gravity of the Ricci flow}
\label{sslapf}

With the improved understanding of the relevant aspects of relativistic and nonrelativistic diffeomorphisms developed in the previous section, we are now ready to examine the supersymmetric theory of gravity with nonprojectable lapse constructed in \cite{grf}, and see if it comes from a simple BRST gauge fixing of an underlying symmetry acting on the ADM metric variables $g_{ij}, n^i$ and $n$.  First, we need to solve the superfield constraints (\ref{eelcon1}), (\ref{eelcon2}) and (\ref{eelcon3}) of the lapse sector in terms of the component fields.  

\subsection{Solving the constraints}
\label{ssconstr}

The process of solving the constraints for the lapse sector superfields follows the steps parallel to the steps we took in solving for the components of the shift superfield $N^i, S^i$ and $\bar S^i$ in Section~\ref{ssdifni}.

First, we solve (\ref{eelcon1}) and find
\bea
\Theta&=&-\nu+\theta(-\nu\dot\nu-\sigma^k\p_k\nu+n^k\nu\p_k\nu)-\bar\theta z\nonumber\\
&&\quad{}+\theta\bar\theta(-\dot\nu-z\dot\nu+\nu\dot z-Y^k\p_k\nu+\sigma^k\p_kz
+\chi^k\nu\p_k\nu+n^kz\p_k\nu-n^k\nu\p_kz),
\eea
and then solving (\ref{eelcon2}) gives
\bea
\bar\Theta&=&-\bar\nu-\theta w+\bar\theta(-\bar\nu\dot{\bar\nu}-\bar\sigma^k\p_k\bar\nu+n^k\bar\nu\p_k\bar\nu)\nonumber\\
&&\quad{}+\theta\bar\theta(w\dot{\bar\nu}-\bar\nu\dot w+\bar\sigma^i\p_i\bar\sigma^k\p_k\bar\nu-\bar\sigma^k\p_kw-\psi^k\bar\nu\p_k\bar\nu-n^kw\p_k\bar\nu+n^k\bar\nu\p_kw).
\eea
Here we must remember that $Y^i$ and $\chi^i$ appearing in the top component of $\Theta$ are composite fields, expressed in terms of the independent component fields $n^i, X^i, \sigma^i$ and $\bar\sigma^i$ of the shift sector as
\bea
Y^i&=&-n^i-X^i+\sigma^k\p_k\bar\sigma^i+\bar\sigma^k\p_k\sigma^i, \\
\chi^i&=&\dot{\bar\sigma}^i+\bar\sigma^k\p_k n^i-\p_k\bar\sigma^i n^k.
\eea

Then we can simply evaluate the components of the $E$ superfield by evaluating the right-hand side of (\ref{eelcon3}).  For the inverse lapse $e$ we get
\be
\label{eecomplapse}
e=1+w+z-\nu(\dot{\bar\nu}-n^k\p_k\bar\nu)-\bar\nu(\dot\nu-n^k\p_k\nu)-\sigma^k\p_k\bar\nu-\bar\sigma^k\p_k\nu.
\ee
This relation is very similar to (\ref{eecompshift}), and we will handle it in the same way:  Our independent bosonic component fields will be the inverse lapse $e$, plus the field $w$ which satisfies $Q\,w=0$ and therefore is our candidate auxiliary field of the antighost-auxiliary multiplet.  Thus, (\ref{eecomplapse}) serves to express $z$ as a composite field in terms of the independent bosonic and fermionic component fields $e, w,\nu$ and $\bar\nu$.

The two fermionic components of the inverse lapse superfield $E$ can also be easily evaluated by applying (\ref{eelcon3}), 
\bea
\psi&=&\nu(\dot e-n_k\p_ke)-(\dot\nu-n^k\p_k\nu)e+\sigma^k\p_ke,\\
\chi&=&\bar\nu(\dot e-n_k\p_ke)-(\dot{\bar\nu}-n^k\p_k\bar\nu)e+\bar\sigma^k\p_ke,
\eea
and so can the top bosonic component $B$, although its explicit expression is not particularly illuminating and will not be useful in the rest of the paper.  

\subsection{BRST interpretation of the component fields and gauge symmetries}

Here we review all the $Q$ multiplets, in the component language, which we obtained in Sections~\ref{ssdifni} and \ref{ssconstr} above by solving the superfield constraints for the lapse and shift superfields derived in \cite{grf}.  We can then check that these multiplets of the nonrelativistic gravity from \cite{grf} can indeed be interpreted as BRST multiplets of a one-step gauge-fixing of a non-redundant gauge symmetry which acts by a combination of topological transformations and spacetime diffeomorphisms.  

First, we have the original $Q$ multiplet containing the spatial metric,
\be
Q\,g_{ij}=\psi_{ij},\qquad Q\,\psi_{ij}=0.
\ee
$\psi_{ij}$ is indeed the topological BRST ghost, associated with the topological symmetry generated by $\delta g_{ij}=\hat f_{ij}$.  The spatial metric supefield $G_{ij}$ also contains the multiplet
\be
Q\,\chi_{ij}=-B_{ij}\qquad Q\,B_{ij}=0.
\label{eeauxg}
\ee
Here $\chi_{ij}$ is the BRST antighost, and $-B_{ij}$ its associated auxiliary field.  

Next we summarize the structure of the $Q$ multiplets in the lapse and shift sectors.  The component solutions of the superspace constraints that we found in Section~\ref{ssdifni} show that the shift vector $n^i$ forms a multiplet with $\sigma^i$,
\bea
Q\,n^i&=&\dot\sigma^i+\sigma^k\p_kn^i-\p_k\sigma^in^k,\\
Q\,\sigma^i&=&\sigma^k\p_k\sigma^i.
\eea
This is indeed the standard BRST multiplet for gauge fixing spatial diffeomorphisms $\hat\xi^i$ acting on the shift vector, when we identify $\sigma^i$ as the ghost associated with the diffeomorphism generator $\hat\xi^i$.  The shift sector also contains the trivial $Q$ multiplet
\be
Q\,\bar\sigma^i=X^i,\qquad Q\,X^i=0.
\label{eeauxni}
\ee
Here $\bar\sigma^i$ is naturally interpreted as the antighost, and $X^i$ its associated auxiliary field.

In the lapse sector, we also find two multiplets, but since the constraints have involved also the shift-vector superfields, the $Q$ transformations of the components are a bit more intricate,
\bea
Q\,e&=&\nu(\dot e-n^k\p_k e)-(\dot\nu-n^k\p_k\nu)e+\sigma^k\p_ke,\\
Q\,\nu&=&\nu(\dot\nu-n^k\p_k\nu)+\sigma^k\p_k\nu,
\eea
as one can see from the component solutions of the lapse-sector constraints found in Section~\ref{ssconstr}.  Happily, on closer inspection (and recalling that $e=1/n$), these happen to be the BRST transformation rules (\ref{eeqtbc2}) associated with the $\hat\zeta$ time-reparametrization symmetry, if we identify $\nu$ as the ghost field $c$.

Thus, we conclude that the $Q$ transformation rules of the multiplets containing $g_{ij}$, $n^i$ and $e$ are exactly the standard BRST tranformation rules (\ref{eeqtbc2}) we obtained in the process of gauge fixing of the spacetime reparametrization and topological symmetries (\ref{eefinalsym}) that we discussed in the previous section!  In addition, among the components of the lapse-sector superfields we have also found the antighost-auxiliary multiplet 
\be
Q\,\bar\nu=w,\qquad Q\,w=0. 
\ee
Together with the antighost-auxiliary multiplets (\ref{eeauxg}) and (\ref{eeauxni}) found in the spatial metric sector and the shift sector, these multiplets are ready to be used in the standard way for implementing the BRST gauge fixing conditions.  

We see that the multiplet structure of the topological nonrelativistic gravity constructed in \cite{grf} precisely reproduces the standard BRST multiplets associated with our gauge symmetries (\ref{eefinalsym}), and gives the correct number of the antighost-auxiliary multiplets needed for the gauge fixing.  The supercharge $Q$ from \cite{grf} is precisely the standard BRST charge associated with the symmetries (\ref{eefinalsym}).  Since these symmetries are non-redundant, the BRST construction closes after the first step, and no ghost-for-ghost fields are needed.   This clarifies the meaning of the two-step superspace construction of our topological gravity in \cite{grf}, and answers fully all the questions about this theory that we raised in the Introduction.

\subsection{Action functionals and Wess-Zumino gauge}

Having clarified the structure of the BRST multiplets, and in particular having understood that the supercharge $Q$ is the standard BRST charge associated with our gauge symmetries (\ref{eefinalsym}), it is time to use the BRST machinery and construct the appropriate gauge-fixed action $S$; or, more precisely, a family of such actions, parametrized by various coupling constants.

The only action $S_0(g_{ij},n^i,n)$ that is invariant under our gauge symmetries is zero, or more precisely, a sum of available topological invariants (whose precise list can depend on the spacetime dimension).  No local dynamics is induced by this gauge-invariant action $S_0$; as usual in topological field theories of the cohomological type \cite{ewcoho}, the entire dynamics will be generated by the judicious choice of the gauge-fixing part of the action, which should take the form
\be
S=\int dt\,d^Dx\,\left\{Q,\Psi\right\},
\label{eegff}
\ee
with the gauge-fixing fermion $\Psi$ being a local functional built out of the available BRST multiplets.  Since we insist on the $\CN=2$ superalgebra (\ref{eesusy}) with supercharges $Q$ and $\bar Q$, the most efficient method for constructing an action consistent with this extended BRST supersymmetry is the superspace approach used in \cite{grf}.  The action will thus be written as
\be
S=\int dt\,d^Dx\,d^2\theta\, \CL,
\label{eegflag}
\ee
where the superspace Lagrangian $\CL$ is a local functional of all the superfields of \cite{grf}.  The superspace geometry then automatically implies that every such superspace action is of the form (\ref{eegff}) for some $\Psi$.

We are dealing with a nonrelativistic theory, which has two \textit{a priori} unrelated scales $L$ and $T$ as discussed in Section~\ref{ssscales}, so the next step is to decide which values of the dynamical exponent $z$ are appropriate for our goals.  Since we wish to make contact with the Ricci flow equations, which happen to select the value of $z=2$, we will be interested in writing down the action that respects this $z=2$ scaling at short distances, and contains the minimal number of time derivatives on the component fields.  As we did in \cite{grf}, it is useful to write such an action in superspace as a sum of two terms,
\be
S=\frac{1}{\kappa^2}(S_K-S_\CW),
\ee
where the ``kinetic'' term $S_K$ contains at least one of the supertime derivatives $\RD$, $\barRD$, $\p_t$, while the ``superpotential'' term $S_\CW$ contains no such supertime derivatives.  As we showed in \cite{grf}, in order the achieve the desired $z=2$ scaling at short distances, $S_K$ will contain terms with one $\RD$ and one $\barRD$, while the superpotential $S_\CW$ will contain terms with up to two spatial derivatives.  Interestingly, this two-derivative superpotential is essentially playing the role of Perelman's $\CF$-functional in our theory.  

Finally, we need to decide whether we wish to keep any of the underlying gauge symmetries unfixed, or whether we prefer the fully gauge-fixed version of the theory.  The appropriate choice may be different depending on which of the three symmetries generated by $\hat f_{ij}$, $\hat\xi^i$ and $\hat\zeta$ we consider.  Keeping some of the gauge symmetries unfixed would lead to the so-called ``equivariant'' theory:  Roughly speaking, the BRST cohomology is then defined as the cohomology of the BRST charge on objects that are gauge invariant under the unfixed symmetry.  This is the strategy often followed for example in topological Yang-Mills gauge theories in four dimensions, where only the topological symmetry is gauge-fixed while the unfixed ``equivariant'' symmetry consists of the ordinary Yang-Mills gauge transformations.

Does it make sense to consider a similar equivariant theory in the context of our topological nonrelativistic gravity of the Ricci flow?  In Section~4.4 of \cite{grf}, we constructed an action which was invariant under the (supersymmetric extension of) the foliation-preserving spacetime diffeomorphisms.  Going to Wess-Zumino gauge as discussed in \cite{grf}, the resulting theory is still equivariant with respect to the bosonic foliation-preserving spacetime diffeomorphisms, generated by spacetime-dependent spatial diffeomorphisms $\xi^i(t,x^k)$ and projectable time reparametrizations $f(t)$.  This unfixed gauge symmetry is strictly larger that the gauge symmetries exhibited by Perelman's Ricci flow.

In the present context, to achieve a closer connection to the Ricci flow, it makes more sense to consider a hybrid theory, in which the gauge-fixing fermion $\Psi$ in (\ref{eegff}) (or the Lagrangian $\CL$ in (\ref{eegflag})) is chosen such that not only the topological symmetries $\hat f_{ij}$ but also the time-reparametrization gauge symmetries $\hat\zeta$ are fully gauge-fixed.  On the other hand, it is meaningful to treat the spatial diffeomorphisms generated by $\xi^i$ as an unfixed, equivariant symmetry:  This symmetry can be gauge-fixed later, as outlined in Section~3.5 of \cite{grf}.  Several gauge choices for the spatial diffeomorphism gauge symmetry -- which we referred to as Perelman gauge, Hamilton gauge, and DeTurck gauge \cite{grf} -- are naturally available, and they match standard manipulations known in the mathematical theory of the Ricci flow.

In the superfield language of \cite{grf}, keeping the bosonic symmetries generated by $\xi^i$ unfixed is accomplished by first choosing $\CL$ that is invariant under the superymmetric generalization of the spatial diffeomorphism, and then going to Wess-Zumino gauge, setting the shift superfield equal to its lowest, bosonic component:
\be
N^i=n^i.
\ee
In the language of component fields, this amounts to setting the ghost field $\sigma^i$, as well as the antighost $\bar\sigma^i$ and its auxiliary $X^i$ all equal to zero.  Indeed, this makes sense:  By throwing away the antighost and the auxiliary, we prevent ourselves from gauge-fixing the associated bosonic symmetry generated by $\xi^i$.  In this Wess-Zumino gauge, the action can then be gauge invariant under the remaining bosonic symmetry $\xi^i$.

In contrast, it would not be very interesting to treat the nonprojectable symmetry of
time reparametrizations $\hat\zeta$ equivariantly:  As we pointed out in (\ref{eeaultra}), if we insist on the unfixed \textit{nonprojectable} bosonic symmetry $\hat\zeta(t,x^i)$, the only lowest-derivative Lagrangians that respect this symmetry are ultralocal in space, with the superpotential given by just the cosmological constant term, and leading to a rather trivial theory and no $z=2$ scaling.  

As we shall see in our forthcoming paper \cite{prf}, in which we establish the precise contact with Perelman's equations of the Ricci flow, it will indeed be vital to use this hybrid strategy:  Gauge-fixing the topological symmetries in the $g_{ij}$ and $n$ sector, while keeping the spatial diffeomorphism symmetry manifest.

\subsection{Dual gauge symmetries}

As an aside remark, we note the following interesting curiosity:  Our theory has an alternative dual interpretation.  Indeed, it is possible to interpret the same component fields studied above as having originated from the gauge fixing of a \textit{dual} copy of spacetime diffeomorphisms, with the role of the BRST charge in this dual picture played by the original anti-BRST charge $\bar Q$.  This is a consequence of the extended $\CN=2$ supersymmetry, together with the ``balanced'' property of the theory, which implies a symmetry between the ghosts and antighosts.  

The existence of such a dual interpretation of the theory can be directly verified by taking a closer look at the $\bar Q$ transformations of the component fields.  In the shift sector, we have
\bea
\bar Q\,n^i&=&\dot{\bar\sigma}{}^i-n^k\p_k\bar\sigma^i+\bar\sigma^k\p_k\bar\sigma^i,\\
\bar Q\,\bar\sigma^i&=&\bar\sigma^k\p_k\bar\sigma^i.
\eea
This is indeed the BRST structure obtained by gauge fixing spatial diffeomorphisms, with the antighost $\bar\sigma^i$ of the original intepretation now playing the role of the ghost.  Similarly, in the lapse sector, we find
\bea
\bar Q\,e&=&\bar\nu(\dot e-n^k\p_ke)-(\dot{\bar\nu}-n^k\p_k\bar\nu)e+\bar\sigma^k\p_ke,\\
\bar Q\,\bar\nu&=&\bar\nu(\dot{\bar\nu}-n^k\p_k\bar\nu)+\bar\sigma^k\p_k\bar\nu.
\eea
Again, this is the BRST multiplet obtained from our ultralocal time reparametrization gauge symmetry, with the original antighost $\bar\nu$ now playing the role of the ghost associated with the dual symmetry.

Similarly, the original ghosts and auxiliary fields of $Q$ give rise to the standard antighost-auxiliary multiplets of $\bar Q$ in the dual interpretation.  First, in the shift sector one finds 
\be
\bar Q\,\sigma^i=Y^i,\qquad \bar Q\,Y^i=0.
\ee
Note that the correct identification of the auxiliary field requires that we solve (\ref{eecompshift}) in a different way than we did in (\ref{eeysol}), now expressing $X^i$ in terms of the independent components $n^i$, $Y^i$, $\sigma^i$ and  $\bar\sigma^i$:
\be
X^i=-n^i-Y^i+\sigma^k\p_k\bar\sigma^i+\bar\sigma^k\p_k\sigma^i.
\ee
Thus, the auxiliaries $X^i$ and $Y^i$ of the two dual interpretations are related by a nonlinear field redefinition.  

Similarly, the antighost-auxiliary $\bar Q$ multiplet in the lapse sector is found to be
\be
\bar Q\,\nu=z,\qquad \bar Q\,z=0.
\ee
This also requires that we solve (\ref{eecomplapse}) differently than how we solved it in Section~\ref{ssconstr}, with $w$ expressed in terms of the independent component fields that now include $z$:
\be
w=-1+e-z+\nu(\dot{\bar\nu}-n^k\p_k\bar\nu)+\bar\nu(\dot\nu-n^k\p_k\nu)+\sigma^k\p_k\bar\nu+\bar\sigma^k\p_k\nu.
\ee
As in the shift sector, the auxiliaries $w$ and $z$ in the two dual interpretations are related by this nonlinear redefinition.  The role of the ghosts and antighosts is simply exchanged between the two dual pictures.  

\section{Conclusions}

In this paper, we have analyzed the ingredients of topological nonrelativistic gravity associated with the Ricci flow, presented in \cite{grf}.  In particular, we substantially clarified the structure of the underlying gauge symmetries of this theory, especially in the time sector.  The construction of \cite{grf} used a two-step procedure, starting with the rigid $\CN=2$ nonrelativistic BRST superspace, and then gauging the symmetries of foliation-preserving spacetime diffeomorphisms.  Here we have demonstrated that this theory can be understood in the more traditional way, as a standard one-step BRST gauge fixing of a theory whose dynamical fields are simply the ADM variables of bosonic gravity describing the spatial metric $g_{ij}$, the shift vector $n^i$ and the lapse function $n$, and whose gauge symmetries consist of the topological symmetries acting on the spatial metric, combined with the ``ultralocal'' nonrelativistic limit of spacetime diffeomorphisms.  These gauge symmetries appear most naturally in their ``shifted'' form (\ref{eefinalsym}).  These findings explain the origin of the ingredients obtained in the superspace formulation in \cite{grf}, and the origin of the superfield constraints derived in \cite{grf} from geometric superspace arguments.  In particular, since the underlying gauge symmetries of the one-step construction presented in this paper are non-redundant, it is clear why no ghost-for-ghost fields were needed in the two-step construction in \cite{grf}.

It is now also clear that the theory is a cohomological quantum field theory of the standard  type \cite{ewcoho}, with the supercharge $Q$ from \cite{grf} exactly playing the role of the standard BRST charge in the full theory.  Moreover, we have also shown why this theory is topological, in the sense of containing no local propagating excitations:  Since all of our gauge symmetries, including the time reparametrizations, are nonprojectable, the number of gauge symmetries matches locally the number of dynamical field components, and the number of local propagating degrees of freedom is zero.  We have also pointed out an intriguing dual interpretation of this theory, as having originated from the gauge fixing of a dual copy of spacetime diffeomorphisms, and with the second supercharge $\bar Q$ now playing the role of the BRST charge.

With this improved understanding of the supermultiplets and the underlying gauge symmetries in the topological quantum gravity of the Ricci flow, it is now possible to study the precise relation \cite{prf} between this topological quantum theory and the mathematical theory of Perelman's Ricci flow, which we started uncovering in \cite{grf}.  

\acknowledgments This work has been supported by NSF Grant PHY-1820912.

%%%%%%%%%%%%%%%%%%%%%%%%%%%%%%%%%%%%%%%%%%%%%%%%%%%%%%%%%%%%%%%%%%%%%%%%%%%%%%%
\bibliographystyle{JHEP}
\bibliography{grf}
\end{document}